\documentclass[11pt]{article}
\usepackage{amssymb,amsmath,mathrsfs,enumerate}
\usepackage{graphicx,rotate,multicol}
\usepackage[margin=10pt,labelfont=bf]{caption}
\usepackage{cite}
\usepackage{braket}
\usepackage[utf8]{inputenc}
\usepackage[colorlinks=true,
            linkcolor=red,
            urlcolor=blue,
        citecolor=blue]{hyperref}

\usepackage{color}
\usepackage{lineno}
\long\def\rpl#1!!!#2!!!{\color[rgb]{.7,0,0}{#1} \color{blue}{#2} \color{black}}

\let\tilde=\widetilde

\let\bar=\overline

\def \order(#1){{\mathcal O} \left(#1 \right)}

\def\cd{{\cal D}}
\def\cu{{\cal U}}

\evensidemargin=\oddsidemargin
\def\Eqn#1{Eq.\ (\ref{#1})}
\def\Eqs#1#2{Eqs.\ (\ref{#1}) and (\ref{#2})}
\allowdisplaybreaks

\title	{\Large\bf 
2HDM without FCNC: off the beaten tracks
}

\author {\sf Dipankar  Das\footnote{ddphy@caluniv.ac.in} \\[10pt]
\small\em Department of Physics, University of Calcutta, 92
Acharya Prafulla Chandra Road, Kolkata 700009, India\\ 
 }

\date{}

\begin{document}


\maketitle	

\begin{abstract}
We propose an alternative method of constructing two Higgs-doublet models free from
scalar mediated FCNC couplings at the tree-level. In a toy scenario, we have
presented semi realistic textures for the Yukawa matrices, which can reproduce
the approximate flavor structure in the quark sector. Presence of flavor diagonal
but nonuniversal Yukawa couplings emerges as a distinguishing feature of such
models.
\end{abstract}

\bigskip

\paragraph{Introduction:}
After the discovery of a Higgs-like particle at the Large Hadron Collider~(LHC)\cite{Aad:2012tfa,
Chatrchyan:2012xdj}, it is now time
to settle whether this new particle is the Higgs scalar as predicted by the Standard Model~(SM),
or it is a Higgs-like particle stemming from a more elaborate construction beyond the SM~(BSM).
In view of this, recent years have seen a growing interest for BSM scenarios with an extended Higgs
sector, where the scalar boson observed at the LHC is only the first to appear in a series of many
others to follow. Two Higgs-doublet models~(2HDMs)\cite{Branco:2011iw,Bhattacharyya:2015nca}
constitute one of the simplest examples of
this category and therefore have received a lot of attention  in recent times.

2HDMs extend the scalar potential of the SM by adding an extra Higgs-doublet.
Consequently, we will now have two Yukawa matrices
for each type of fermions and diagonalization of the fermion mass matrix will not guarantee,
in general, the simultaneous diagonalization of the Yukawa matrices. Thus, there will be 
flavor changing neutral currents (FCNC), at the tree level, mediated by the neutral scalars.
However, it has been shown by Glashow and Weinberg\cite{Glashow:1976nt} and independently 
by Paschos\cite{Paschos:1976ay} that absence of tree-level FCNC can be
ensured if suitable arrangements are made such that fermions of a particular charge receive
their masses from a single scalar doublet ($\phi_1$ or $\phi_2$). Usually, a $Z_2$ symmetry, 
under which $\phi_1\to \phi_1$ and $\phi_2\to -\phi_2$, is employed to accomplish this. Proper assignments of the 
$Z_2$ charges to different fermions then achieves the purpose.

To make things explicit, let us write down the Yukawa part of a 2HDM Lagrangian as follows:
\begin{eqnarray}
\label{e:yuklag}
\mathscr{L}_Y = - \sum_{k =
  1}^2\left[\bar{\mathbf Q}_L\Gamma_k \phi_k \mathbf d_{R} +
  \bar{\mathbf Q}_L\Delta_k \tilde{\phi}_k \mathbf u_{R}
  +\bar{\mathbf L}_L\Sigma_k \phi_k {\mathbf e}_{R}\right] +
        {\rm h.c.},  
\end{eqnarray}
where ${\mathbf Q}_L$ and ${\mathbf L}_L$ denote the quark and lepton doublets respectively and
$\mathbf u_{R}$, $\mathbf d_{R}$ and $\mathbf e_{R}$ represent the up, down and charged lepton
singlets respectively. We have also used the standard abbreviation $\tilde\phi_k = i \sigma_2
\phi_k^*$, where $\sigma_2$ is the second Pauli matrix. Note that we have suppressed the 
flavor indices in \Eqn{e:yuklag}. Therefore, $\Delta_k$,
$\Gamma_k$ and $\Sigma_k$ stand for $3\times 3$ Yukawa matrices in the up, down and charged
lepton sectors respectively. We have also assumed that the neutrinos are massless.

From \Eqn{e:yuklag} one can write the mass matrix for the down type quarks, for example, as
follows:
\begin{eqnarray}
\label{e:Md}
M_d = \Gamma_1 \braket{\phi_1} +\Gamma_2 \braket{\phi_2} \,,
\end{eqnarray}
where $\braket{\phi_k}=v_k/\sqrt{2}$ denotes the vacuum expectation value (VEV) of $\phi_k$. Since
$\Gamma_1$ and $\Gamma_2$, in principle, can be arbitrary, there is no reason for them
to be simultaneously diagonal once $M_d$ is diagonalized
using a biunitary transformation. Therefore, in general, there
will be Higgs mediated FCNC at the tree level. Following the Glashow-Weinberg-Paschos (GWP)
prescription, if either $\Gamma_1$ or $\Gamma_2$ vanishes, then the mass matrix becomes
proportional to the Yukawa matrix, just as in the SM, and tree-level FCNC in the down quark
sector can be avoided altogether. The same method can be applied to remove tree-level
FCNC from the up quark and the charged lepton sectors too. The GWP prescription can be
attributed to a $Z_2$ symmetry which prevails in the Yukawa Lagrangian. Consider, for
example, the case with $\Gamma_2=\Delta_2=\Sigma_2=0$ which implies that a $Z_2$
transformation under which only $\phi_2$ is odd, keeps the Yukawa Lagrangian invariant.
This $Z_2$ symmetry, when extended to the full Lagrangian, also prevents the corresponding Yukawa couplings from getting generated via quantum corrections.
Thus the GWP prescription, by design, offers a natural solution to the FCNC problem.

On the other hand, Pich and Tuzon\cite{Pich:2009sp} suggested that an alternative way
to avoid tree-level FCNC will be to make the mass matrix proportional to the Yukawa
matrix by assuming the two Yukawa matrices for fermions of a particular charge to be
proportional to each other. These types of 2HDMs are called aligned 2HDMs (A2HDMs).
However, the
Yukawa alignment imposed at a certain energy scale does not guarantee, in general,
that the alignment will be maintained at a different scale too\cite{Ferreira:2010xe,Botella:2015yfa}. 
But it has been shown that
the FCNCs generated due to such misalignments are expected to be small\cite{Braeuninger:2010td} because
the Yukawa aligned 2HDMs belong to a general category of models with minimal
flavor violation (MFV)\cite{Buras:2010mh}.

\paragraph{An alternative lifestyle:}
However, for two Yukawa matrices to be simultaneously diagonalizable, it is not
necessary to assume one of them to be zero or they are proportional to each other.
It has been shown\cite{Williamson,GIBSON197445} that
if $M_1$ and $M_2$ are two complex square matrices then there exist unitary matrices
$U_1$ and $U_2$ such that both $U_1^\dagger M_1 U_2$ and $U_1^\dagger M_2 U_2$
are diagonal if and only if both $M_1 M_2^\dagger$ and $M_2^\dagger M_1$
are normal matrices\footnote{Additionally, if we also impose the reality of the diagonal elements, then
both $M_1 M_2^\dagger$ and $M_2^\dagger M_1$ need to be hermitian matrices\cite{Eckart}.}.

Keeping in mind the near block diagonal structure of the CKM matrix in the
Wolfenstein parametrization\cite{Wolfenstein:1983yz}, we take, for instance\footnote{
Notice that it is not necessary to assume $(\Gamma_2)_{33}=0$ in \Eqn{e:g1g2}. But as we will
see later, this assumption makes it easier to motivate the mass matrix from
symmetry.}
\begin{eqnarray}
\label{e:g1g2}
\Gamma_1 = \begin{pmatrix}
b_1 & 0 & 0 \\
0  & b_1 & 0 \\
0  & 0   &  b_3
\end{pmatrix} \,,
&&
\Gamma_2 = \begin{pmatrix}
b^{'}_2 & b_2 & 0 \\
b_2  & -b^{'}_2 & 0 \\
0  & 0   &  0
\end{pmatrix} \,.
\end{eqnarray}
These Yukawa matrices will lead to the following mass matrix in the down quark
sector:
\begin{eqnarray}
\label{e:md}
M_d = \frac{1}{\sqrt{2}}\left(\Gamma_1v_1+\Gamma_2v_2\right) =
\frac{1}{\sqrt{2}} \begin{pmatrix}
b_1 v_1 + b^{'}_2 v_2 & b_2 v_2 & 0 \\
b_2 v_2  & b_1v_1 -b^{'}_2 v_2 & 0 \\
0  & 0   &  b_3 v_1
\end{pmatrix} \,.
\end{eqnarray}
We will assume that the Yukawa couplings are real. Under this assumption, $\Gamma_1$
and $\Gamma_2$ become mutually commuting hermitian matrices and therefore they can
be diagonalized simultaneously. As a consequence, there should be no FCNC at the
tree-level mediated by the neutral scalars.

To verify this assertion explicitly, we note that $M_d$ can be diagonalized as follows:
\begin{eqnarray}
\label{e:dLR}
D_d = \cd_L^\dagger M_d \cd_R = {\rm diag}\left\{m_d,m_s,m_b\right\} \,,
\end{eqnarray}
where,
\begin{eqnarray}
\label{e:uni}
\cd_L =\cd_R =
\begin{pmatrix}
\cos\theta_d & \sin\theta_d & 0 \\
-\sin\theta_d  & \cos\theta_d & 0 \\
0  & 0   &  1
\end{pmatrix} \,.
\end{eqnarray}
Then we have the following relations:
\begin{subequations}
\label{e:thd}
\begin{eqnarray}
\label{e:11}
m_d\cos^2\theta_d+ m_s\sin^2\theta_d &=& \left(b_1 v_1 +b^{'}_2 v_2\right)/\sqrt{2} \,, \\
\label{e:22}
m_d\sin^2\theta_d+ m_s\cos^2\theta_d &=& \left(b_1 v_1 -b^{'}_2 v_2\right)/\sqrt{2} \,, \\
\left(m_d-m_s\right)\sin\theta_d \cos\theta_d &=& b_2v_2/\sqrt{2} \,, \\
\tan2\theta_d &=& \frac{b_2}{b^{'}_2} \,.
\end{eqnarray}
\end{subequations}
Similarly we can define unitary matrices $\cu_L$ and $\cu_R$ which diagonalize
the mass matrix in the up quark sector and are characterized by the angle $\theta_u$.
The CKM matrix will then be given by
\begin{eqnarray}
V=\cu_L^\dagger \cd_L =
\begin{pmatrix}
\cos(\theta_d-\theta_u) & \sin\left(\theta_d-\theta_u\right) & 0 \\
-\sin\left(\theta_d-\theta_u\right)  & \cos\left(\theta_d-\theta_u\right) & 0 \\
0  & 0   &  1
\end{pmatrix} \,.
\label{e:CKM}
\end{eqnarray}
Thus, we can identify $\sin\left(\theta_d-\theta_u\right)$ as the Cabibbo mixing
parameter, $\lambda$. We also note that $\Gamma_1$ and $\Gamma_2$ can
be diagonalized simultaneously and using \Eqn{e:thd} we may write,
\begin{eqnarray}
X_1 &=& \cd_L^\dagger \Gamma_1 \cd_R = \frac{\sqrt{2}}{v_1}~ {\rm diag}\left\{\frac{m_d+m_s}{2}, \frac{m_d+m_s}{2}, m_b \right\} \,, \\
X_2 &=& \cd_L^\dagger \Gamma_2 \cd_R = \frac{\sqrt{2}}{v_2}~ {\rm diag}\left\{\frac{m_d-m_s}{2}, \frac{m_s-m_d}{2}, 0 \right\} \,.
\end{eqnarray}
Therefore, there will be no Higgs mediated FCNC at the tree level.

In passing, we note that although \Eqn{e:CKM} gives only an approximate form
of the CKM matrix, it is not very difficult, using a bottom-up approach,
to reconstruct suitable Yukawa textures which are compatible with the full CKM structure. As an
example, let us assume that the Yukawa matrices in the up quark sector,
$\Delta_1$ and $\Delta_2$ are already diagonal, so that we may choose
$\cu_L=\cu_R = 1$. Thus, in this case, the CKM matrix, $V$, becomes identical
to $\cd_L$ which has been defined in \Eqn{e:dLR}. Now, let us also assume
that $\Gamma_1$ and $\Gamma_2$ have been diagonalized simultaneously as
\begin{eqnarray}
\label{e:fullCKM}
	X_1 = V^\dagger \Gamma_1 \cd_R \,, \qquad
	X_2 = V^\dagger \Gamma_2 \cd_R \,,
\end{eqnarray}
so that we may write
\begin{eqnarray}
	X_1 \frac{v_1}{\sqrt{2}} + X_2 \frac{v_2}{\sqrt{2}}
	= {\rm diag}\left\{m_d,m_s,m_b\right\} \,.
\end{eqnarray}
We can easily find nontrivial solution sets for $X_1$ and $X_2$, which satisfy
the relation above. We can then choose any structure for the unitary matrix,
$\cd_R$ and invert \Eqn{e:fullCKM} to obtain a set of solutions for the
Yukawa matrices, $\Gamma_1$ and $\Gamma_2$, which are congenial to the full
CKM structure, yet can ensure the absence of FCNC at the tree-level.
However, one should keep in mind that such a bottom up approach contains
too much arbitrariness in the resulting Yukawa matrices which might be difficult
to motivate from an underlying symmetry.

\paragraph{Impact on flavor universality:}
If we express the scalar doublet, $\phi_k$  as
\begin{eqnarray}
	\phi_k = \begin{pmatrix}
	w_k^+ \\ \frac{h_k+iz_k}{\sqrt{2}}
	\end{pmatrix} \,,
\end{eqnarray}
then it is well known that the combination $h=(v_1h_1+v_2h_2)/v$, where 
$v=\sqrt{v_1^2+v_2^2}\approx 246$~GeV, has SM-like gauge and
Yukawa couplings at the tree-level. The limit where $h$ corresponds to a physical
eigenstate is known as the {\em alignment limit}\cite{Carena:2013ooa,Dev:2014yca,Bhattacharyya:2014oka,Bernon:2015qea} 
which is being increasingly favored
by the LHC Higgs data\cite{Bernon:2014vta}. In the down quark sector, the Yukawa couplings for $h$ and
its orthogonal combination, $H=(v_2h_1-v_1h_2)/v$, are given by
\begin{eqnarray}
\label{e:Yuk}
{\mathscr L}_Y^d = -\frac{h}{v}\bar{\mathbf d}\, D_d\, {\mathbf d} -\frac{H}{v}\bar{\mathbf d}\left(
N_dP_R+N_d^\dagger P_L\right) {\mathbf d} = -\frac{h}{v}\bar{\mathbf d} \, D_d \, {\mathbf d} 
-\frac{H}{v}\bar{\mathbf d}
\, N_d \, {\mathbf d} \,,
\end{eqnarray}
where, ${\mathbf d}=(d,s,b)^T$ and in writing the last step, it has been assumed that $N_d$
is a real diagonal matrix. For our choice of Yukawa textures in \Eqn{e:g1g2}, $N_d$ has the following form:
\begin{eqnarray}
\label{e:Nd}
N_d = {\rm diag}\left\{\frac{m_d+m_s}{2}\tan\beta -\frac{m_d-m_s}{2}\cot\beta, \,
\frac{m_d+m_s}{2}\tan\beta- \frac{m_s-m_d}{2}\cot\beta, \, m_b\tan\beta \right\} \,,
\end{eqnarray}
where $\tan\beta=v_2/v_1$.
Looking at this structure of $N_d$, we can see that the nonstandard scalar, $H$, couples
to the fermions in a flavor diagonal but  nonuniversal manner. This is in sharp
contrast with other flavor conserving 2HDMs {\em e.g.}, the $Z_2$ symmetric 2HDMs
and the A2HDM. To be more precise, we define the following quantity:
\begin{eqnarray}
\label{e:af}
	\alpha_f = \frac{v\, Y_{Hff}}{m_f} \,,
\end{eqnarray}
where $Y_{Hff}$ denotes the diagonal Yukawa coupling of $H$ with the $f\bar{f}$ pair
and $m_f$ represents the mass of the fermion, $f$. In $Z_2$ symmetric 2HDMs and in
A2HDM, the quantity, $\alpha_f$, for fermions of a particular charge, is a constant independent
of the fermion masses. For instance, in the case of type~II 2HDM, $\alpha_f=\tan\beta$
for the down type quarks. From the expression of $N_d$ in \Eqn{e:Nd}, we see that this
is clearly not satisfied for our choice of Yukawa matrices in \Eqn{e:g1g2}. It is in this
sense that we claim that the Yukawa couplings for the nonstandard Higgs bosons are
flavor nonuniversal.

It is worth emphasizing that this nonuniversality is not a special outcome of our
particular choice for the Yukawa matrices in \Eqn{e:g1g2}. Whereas the precise nature
of the flavor nonuniversal couplings depends on the details of the textures for the
Yukawa matrices, the mere presence of nonuniversality is a general artefact of such
constructions. To illustrate our point, we note that the matrix, $N_d$, introduced
in \Eqn{e:Yuk} has the following general form:
\begin{eqnarray}
\label{e:Nd1}
	N_d = \frac{1}{\sqrt{2}}\cd_L^\dagger \left(v_2\Gamma_1- v_1\Gamma_2\right) \cd_R
	=\frac{1}{\sqrt{2}} \left(v_2X_1- v_1X_2\right) \,.
\end{eqnarray}
In writing \Eqn{e:Nd1} we have assumed that $\Gamma_1$ and $\Gamma_2$ have
been diagonalized simultaneously to $X_1$ and $X_2$ respectively. Then following
\Eqn{e:af} we may write,
\begin{eqnarray}
\label{e:aq}
\alpha_q = \frac{(N_d)_{qq}}{m_q} = \frac{v_2(X_1)_{qq} -v_1 (X_2)_{qq}}{
v_1(X_1)_{qq} +v_2 (X_2)_{qq}} \,,
\end{eqnarray}
where the subscript, $q$, stands for a generic down type quark and $m_q$ has
been obtained by diagonalizing \Eqn{e:Md}. Now, if we demand
$\alpha_q =\alpha$ to be a constant for the down type quarks, then we will have,
\begin{eqnarray}
\label{e:align}
X_1 \left(\tan\beta -\alpha\right) = X_2 \left(1+\alpha\tan\beta \right) \,
\end{eqnarray}
which implies that the Yukawa matrices, $\Gamma_1$ and $\Gamma_2$, should be
proportional to each other. Note that, conventional $Z_2$ symmetric 2HDMs constitute
very special cases of the A2HDM scenario\cite{Ferreira:2010xe}. For instance, if we have either $\Gamma_1=0$
or $\Gamma_2=0$ then, using \Eqn{e:align}, we can recover $\alpha=-\cot\beta$ or
$\alpha=\tan\beta$ as expected for type~I and type~II 2HDMs respectively\cite{Branco:2011iw}.

Thus we have shown that for the nonstandard scalars to have flavor universal Yukawa 
couplings, the 2HDM must either correspond to a conventional 2HDM with $Z_2$
symmetry or to the A2HDM. Any other way of constructing 2HDM without FCNC will inevitably
lead to flavor nonuniversal couplings which can serve as a distinguishing feature
of such constructions.

\paragraph{A symmetry origin:}
At this stage, it is reasonable to ask whether the Yukawa structures presented in
\Eqn{e:g1g2} will be stable under quantum corrections. Evidently, the answer will
be affirmative if we can find a symmetry that can give rise to such textures. As it happens,
we can motivate the Yukawa structures of \Eqn{e:g1g2} within a 2HDM framework using 
an approximate $Z_3\times Z_2$ flavor symmetry. First we assign the following
transformation properties to the quark and scalar fields under $Z_3$\footnote{
This will lead to an accidental $U(1)$ symmetry in the scalar potential, under
which $\phi_1\to e^{i\theta}\phi_1 \,; \phi_2\to \phi_2$. Consequently, a massless
pseudoscalar should appear after the SSB. This can be prevented by introducing
a bilinear term of the form $\{m_{12}^2 (\phi_1^\dagger \phi_2) + {\rm h.c.}\}$
to the scalar potential, which breaks the symmetry softly. Details of the scalar
spectrum for such a model can be found in Ref.~\cite{Bhattacharyya:2013rya}.}:
\begin{eqnarray}
	Q_{L3} \to \omega \, Q_{L3} \,, \qquad u_{R3} \to \omega^2 u_{R3} \,,
	\qquad \phi_1 \to \omega \, \phi_1 \,,
\end{eqnarray}
where, $Q_{Lk}$ $(k=1,2,3)$ stands for the left handed $SU(2)$ doublet of quarks
in the $k$-th generation and
similarly for the right handed quark singlets in the up and down sectors, and $\phi_{k}$ 
$(k=1,2)$ denotes the $k$-th scalar doublet of $SU(2)$. The scalar doublet, $\phi_2$,
and the other quark fields remain unaffected under $Z_3$.
Under the $Z_2$ part of the
symmetry only $Q_{L3}$, $u_{R3}$, and $d_{R3}$ are assigned odd parities while
the rest are assumed to be even. With these assignments, certain terms such as
$\bar{Q}_{L3}(\Gamma_2)_{3k}\phi_2 d_{Rk}$ $(k=1,2,3)$ will be forbidden in
the Yukawa Lagrangian of \Eqn{e:yuklag}. One can easily verify that, because
of the symmetry, the Yukawa matrices will take the following forms\footnote{
If the Yukawa textures of \Eqn{e:symtex} were exact, then flavor universality
would prevail separately in the first two generations and in the third generation
of fermions.}:
\begin{eqnarray}
\label{e:symtex}
	\Gamma_1, \, \Delta_1 = 
	\begin{pmatrix}
	0 & 0 & 0 \\
	0 & 0 & 0 \\
	0 & 0 & \checkmark
	\end{pmatrix} \,, \qquad
	\Gamma_2, \, \Delta_2 = 
	\begin{pmatrix}
	\checkmark & \checkmark & 0 \\
	\checkmark & \checkmark & 0 \\
	0 & 0 & 0
	\end{pmatrix} \,.	
\end{eqnarray}
Thus, as long as only the block configurations are concerned, the structures
of \Eqn{e:symtex} look very similar to those in \Eqn{e:g1g2} except for the
fact that $b_1=0$ in \Eqn{e:symtex}, which is a consequence of the $Z_3$
part of the symmetry. But adding \Eqs{e:11}{e:22} we can see that $b_1$
is of the order of the Yukawa couplings for the second generation of
quarks, which for $\tan\beta \sim \order(1)$ is quite small. Therefore,
even in the presence of $b_1$ (and its counterpart in the up quark
sector) the Yukawa Lagrangian can be considered to possess an
approximate $Z_3 \times Z_2$ symmetry which is expected to protect
the zeros in \Eqn{e:g1g2} from large quantum corrections.
Moreover, when $N_d$ and $N_u$ (analog of $N_d$ in the up sector) are
diagonal, the Yukawa Lagrangian possesses an approximate
$U(1)_1\times U(1)_2\times U(1)_3$ symmetry\footnote{
	If the CKM matrix were the unit matrix, then, under this symmetry, the
	individual generation numbers would have been conserved in the quark sector.}
which is broken only by the off-diagonal elements of the CKM matrix. Thus,
in addition to the usual loop suppression factor, the FCNC couplings generated by
quantum effects will be moderated further by the off-diagonal CKM elements.

\paragraph{Summary:}
To summarize, we have proposed a novel way of constructing 2HDMs devoid of tree-level
FCNCs mediated by neutral scalars. We have argued how, to have a FCNC free 2HDM, it
is not necessary to assume the vanishing of one of the Yukawa matrices or the
proportionality between the two Yukawa matrices for fermions of a particular charge.
We have shown, with a semi-realistic example in the quark sector, that 
for judiciously chosen nontrivial structures for the Yukawa matrices
one can still prevent FCNCs from appearing at the tree-level. Although the main
idea of this paper is simple in its conception, yet, to our knowledge, this is
the first time that such a possibility has been pointed out explicitly.
Despite the simplicity, the unavoidable presence of flavor diagonal but nonuniversal
Yukawa couplings for the nonstandard scalars is a unique and interesting
outcome which should enable us to experimentally distinguish such constructions
from the existing conventional 2HDMs. In view of the fact that recent
anomalies in the $B$-physics data may call for flavor nonuniversal
couplings\cite{Becirevic:2015asa,Altmannshofer:2017yso,Capdevila:2017bsm}, 
such a 2HDM can open up new possibilities along this direction.

\section*{Appendix}
For the sake of completeness and to give the reader an intuitive feel, we present
here the requirement for simultaneous diagonalization of two complex matrices as
a necessary condition. First we recall that a normal matrix commutes with its
hermitian conjugate.
%
Now, if we assume that $N$ is diagonalizable via a unitary similarity transformation,
{\it i.e.}, $U^\dagger N U=D$ where $D$ is a diagonal matrix, then we have
\begin{eqnarray}
\label{e:1}
NN^\dagger = \left(UDU^\dagger \right) \left(U D^\dagger U^\dagger \right)
= U D D^\dagger U^\dagger = U D^\dagger D U^\dagger =
\left(U D^\dagger U^\dagger \right) \left(U D U^\dagger \right) = N^\dagger N \,,
\end{eqnarray}
which implies that $N$ is a normal matrix. Note that in \Eqn{e:1} we have used
the fact that $D$, being diagonal, commutes with $D^\dagger$.

Conversely, we can also show that if $N$ is a normal matrix then it can be diagonalized
via a unitary similarity transformation. To prove this, we note that $N$ can be decomposed
as:
\begin{eqnarray}
\label{e:n1n2}
N= N_1 + i N_2 \,,
\end{eqnarray}
where $N_1$ and $N_2$ are both hermitian matrices and are given by
\begin{eqnarray}
N_1 = \frac{1}{2}\left(N+N^\dagger \right) \,, ~~~ {\rm and} ~~
N_2 = -\frac{i}{2}\left(N-N^\dagger \right) \,.
\end{eqnarray}
Now, one can check that, if $N$ is a normal matrix, {\it i.e.}, $[N,N^\dagger]=0$,
then we must have, $[N_1,N_2] = 0$.
Thus, $N_1$ and $N_2$, being hermitian matrices, can be diagonalized by the
same unitary similarity transformation, which means, in view of \Eqn{e:n1n2},
that $N$ can be diagonalized via a unitary similarity transformation. Hence, we
have proved that a matrix, $N$, can be diagonalized via a unitary similarity
transformation if and only if $N$ is a normal matrix.

Now, let us suppose that two complex Yukawa matrices, $Y_1$ and $Y_2$, can
be diagonalized simultaneously via the following biunitary transformation:
\begin{eqnarray}
U^\dagger_1 Y_1 U_2 = D_1 \,, ~~~ {\rm and} ~~
U^\dagger_1 Y_2 U_2 = D_2 \,.
\end{eqnarray}
Then we can write,
\begin{eqnarray}
\label{e:d1}
&& \left(U^\dagger_1 Y_1 U_2 \right) \left(U^\dagger_2 Y_2^\dagger U_1 \right)
\equiv U^\dagger_1 \left(Y_1 Y_2^\dagger \right) U_1 = D_1 D_2^\dagger \,, \\
\label{e:d2}
{\rm and}
&& \left(U^\dagger_2 Y_1^\dagger U_1 \right) \left(U^\dagger_1 Y_2 U_2 \right)
\equiv U^\dagger_2 \left(Y_1^\dagger Y_2 \right) U_2 = D_1^\dagger D_2 \,.
\end{eqnarray}
Since the matrices on the right hand sides of \Eqs{e:d1}{e:d2} are diagonal, we
can conclude that $Y_1Y_2^\dagger$ and $Y_1^\dagger Y_2$ must be normal
matrices.


\bibliographystyle{JHEP} 
\bibliography{Yuk.bib}
\end{document}